\documentclass[twocolumn,english,aps,prl,superscriptaddress]{revtex4-1}
\usepackage[latin9]{inputenc}
\usepackage{amsmath}
\usepackage{amssymb}
\usepackage{graphicx}

\makeatletter
\usepackage{blindtext}\usepackage{float}
\usepackage{color}\usepackage{upgreek}
\usepackage{lipsum}

\makeatother

\usepackage{babel}
\begin{document}

\title{ Precession Motion in Levitated Optomechanics}

\author{Muddassar Rashid}
\email{m.rashid@soton.ac.uk}

\selectlanguage{english}%

\affiliation{Department of Physics and Astronomy, University of Southampton, SO17
1BJ, UK}

\author{Marko Toro\v{s}}
\email{m.toros@soton.ac.uk}

\selectlanguage{english}%

\affiliation{Department of Physics and Astronomy, University of Southampton, SO17
1BJ, UK}

\author{Ashley Setter}

\affiliation{Department of Physics and Astronomy, University of Southampton, SO17
1BJ, UK}

\author{Hendrik Ulbricht}
\email{h.ulbricht@soton.ac.uk}

\selectlanguage{english}%

\affiliation{Department of Physics and Astronomy, University of Southampton, SO17
1BJ, UK}
\begin{abstract}
  We investigate experimentally the dynamics of a non-spherical
  levitated nanoparticle in vacuum.  In addition to translation and
  rotation motion, we observe the light torque-induced precession and
  nutation of the trapped particle. We provide a theoretical model,
  which we numerically simulate and from which we derive
  approximate expressions for the motional frequencies. Both, the
  simulation and approximate expressions, we find in good agreement
  with experiments. We measure a torque of
  $1.9 \pm 0.5 \times 10^{-23}$ Nm at $1 \times 10^{-1}$ mbar, with an
  estimated torque sensitivity of $3.6 \pm 1.1 \times 10^{-31}$
  Nm/$\sqrt{\text{Hz}}$ at $1 \times 10^{-7}$ mbar. 
\end{abstract}
\maketitle

{\it Introduction--} A typically levitated optomechanical setup
comprises of a particle, which is trapped using a tightly focussed
laser beam. The trapped particle is best described as a harmonic
oscillator and can exhibit a rich variety of dynamics. For example,
the dynamics can be under-damped or over-damped depending on the
number of collisions controlled by the background gas pressure. The
dynamics can be linear or can show strong non-linearities depending on
the oscillation amplitude, which is controlled by temporal and spatial
modulated external electric, magnetic, and light forces: the dynamics
can be driven or cooled. This tunability of the dynamics has enabled
various studies of the rich physics of this harmonic oscillator, among
them Brownian motion~\cite{Li2010a}, nonlinear dynamics
\cite{Gieseler2013, Fonseca2016}, test of fluctuation theorems and
non-equilibrium physics \cite{Gieseler2014,Gieseler2014b, Hoang2018},
and thermodynamics in the single particle regime
\cite{Millen,Rondin2017,Aranas2017}. The unique properties of the
underlying dynamics and the ability to control levitated optomechanics
demonstrate its immense potential for sensing~\cite{Ranjit2016,
  Hebestreit2017,Hempston2017}, as well as, to address fundamental
questions in physics~\cite{Romero-Isart2011,Bassi2013,bateman2014}.

The centre-of-mass (c.o.m.) translation motion of the trapped
nanoparticle has been studied in quite some detail already. Optical
feedback, cavity-assisted schemes and electric forces have been used
to control the translation and have been ultimately used to cool the
motion to millikelvin temperatures
\cite{Li2011a,Kiesel2013,Vovrosh2016,Setter2017} and below
\cite{vijay2016}, already close to the ground state of the harmonic
oscillator.

In addition, to exhibiting translation motion, trapped particles can
also show rotation ~\cite{Arita2013,Kuhn2015,Rahman2017} and libration
~\cite{Hoang2016} motion. Different light-matter physics has been used
to drive and control the rotation of levitated particles, such as the
polarizability-anisotropy in silicon rods coupling to the polarization
of light~\cite{Kuhn2016}. Rotational frequencies of up to some GHz
\cite{Reimann2018, Ahn2018} have been observed, only limited by the
centrifugal damage threshold of the rotating particle.

Variation of the linear to circular polarization of light gives a
handle to switch between rotation and libration. Libration has been
demonstrated experimentally with trapped nanodiamonds
\cite{Hoang2016}, silicon rods \cite{Kuhn2016} and dumbbells
\cite{Ahn2018}. Like translation, libration motion is described by a
harmonic oscillator model and is therefore a candidate to apply
similar optical techniques for cooling with reasonable promise to
reach a quantum ground state, making libration a stark contender in
the race towards the quantum regime. First proposals discuss the
usefulness of libration to generate macroscopic quantum states such as
angular superpositions \cite{Carlesso2017, Ma2017}. Additionally, both
rotation and libration motion promise unprecedented high levels of
sensitivity~\cite{Hoang2016,Kuhn2017,Stickler2018} for detection of
weak forces such as gravity \cite{Carlesso2017,Carlesso2017a} and
dispersive forces \cite{Manjavacas2017}.

In this letter, we report on the observation of light-induced
precession motion of a non-spherical silica particle compound. We give
a theoretical description of the system and numerically simulate the
model. We identify the mechanical frequencies in the experimental
spectrum: in particular, translation, rotation (spin), precession, and
nutation motion. We investigate the precession by variation of
background gas pressure and of the power of the trapping laser in
agreement with the theoretical model. We discuss the possibilities for
torque sensing applications.
\begin{figure*}[t]
  \centering
  \includegraphics[width=1\linewidth]{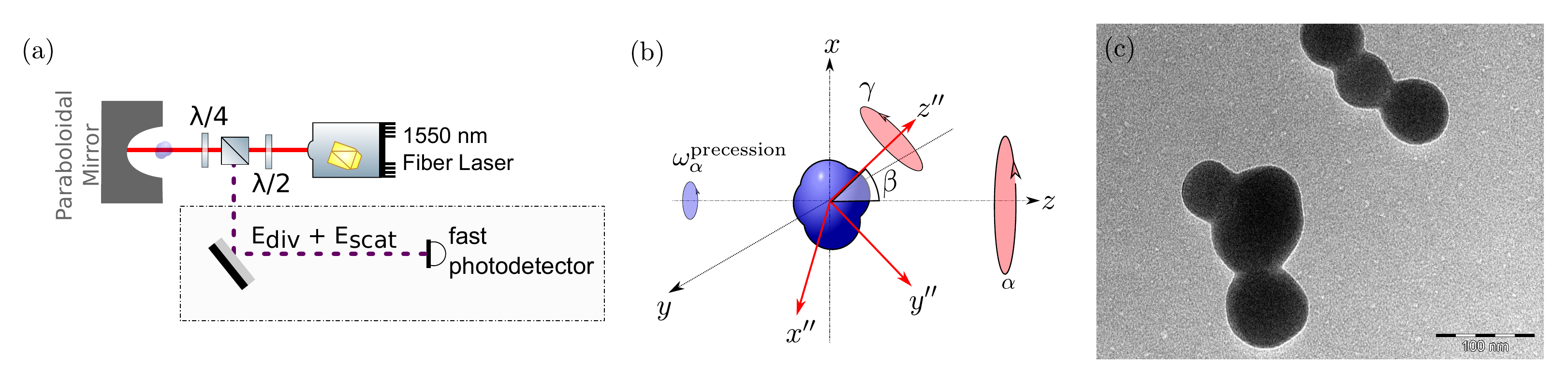}
  \caption{\textbf{The Levitated Optomechanical System:} (a) A 1550 nm
    laser beam is focused by the paraboloidal mirror, the particle is
    trapped in the focus. Once trapped, the scattered light,
    $E_{\text{scat}}$, from the particle is collected and directed
    towards the detection system. The interference between the
    scattered and the diverging electric fields, $E_{\text{div}}$, is
    used to detect the motion of the particle. (b) Paremetrization: A
    non-spherical particle that is trapped by a laser field
    propagating in the $z$-direction.  The laboratory axis are denoted
    by $x$, $y$, $z$, while the body-frame axis are denoted by the
    $x''$, $y''$, $z''$. The relation between the two frames is
    parametrized by the Euler angles $\alpha$, $\beta$ and $\gamma$ in
    $z$-$y'$-$z''$ convention. $\alpha$ denotes the angle of rotation
    about the laboratory $z$-axis (from $x$ towards $y$). $\beta$ is
    the angle between the laboratory $z$-axis and the body $z''$-axis
    (rotated about the $y'$-axis, i.e. the $y$-axis after it has been
    rotated by $\alpha$ about the $z$-axis; from $z$ towards
    $x$). $\gamma$ denotes the angle of rotation about the body frame
    $z''$-axis (from $x''$ towards $y''$). (c) TEM image of compound
    silica particle solution in water, few weeks after
    preparation. The original particle solution is made of 50 nm
    radius silica spheres (Corpuscular Inc.).}
  \label{fig:theory}
\end{figure*}

{\it Theoretical model--} We consider an anisotropic polarizable
particle, which is optically trapped by an elliptically polarized
Gaussian laser beam. Part of the scattered photons are collected and
directed, using optical elements, towards a single photodetector. The
scattered light is mixed with a local oscillator to obtain the direct
homodyne photo-current $I_{\text{exp}}$, see Fig. \ref{fig:theory}(a)
~\cite{Rashid2017}. This experimental situation has been analyzed
theoretically using a quantum model~\cite{Toros2018t}, as well as
numerically, using an approximate classical
model~\cite{Toros2018s}. We now summarize the dynamics referring to
the latter, where we assume that we are at relatively high pressure,
such that we can neglect photon-recoil heating terms.

The dynamics can be described in a twelve dimensional phase space of
the following classical variables: center of mass position
$\boldsymbol{r}=(x,y,z)^{\top}$, center of mass conjugate momentum
$\boldsymbol{p}=(p_{x},p_{y},p_{z})^{\top}$, angle
$\boldsymbol{\phi}=(\alpha,\beta,\gamma)^{\top}$, and angle conjugate
momentum
$\boldsymbol{\pi}=(\pi_{\alpha},\pi_{\beta},\pi_{\gamma})^{\top}$,
where the angles are defined in the Euler $z$-$y'$-$z''$ convention
(see Fig. \ref{fig:theory}(b)). In particular, the dynamics is given
by (in It\^{o} form):
\begin{alignat}{1}
d\boldsymbol{r}= & -\boldsymbol{\partial_{p}}H_{\text{free}}dt\label{eq:dr}\\
d\boldsymbol{p}= & -\boldsymbol{\partial_{r}}H_{\text{grad}}dt+d\boldsymbol{p}_{\text{scatt}}+d\boldsymbol{\boldsymbol{p}}_{\text{coll}}\label{eq:dp}\\
d\boldsymbol{\boldsymbol{\phi}}= & \boldsymbol{\partial_{\pi}}(H_{\text{free}}+H_{\text{grad}})dt\label{eq:dphi}\\
d\boldsymbol{\boldsymbol{\pi}}= & -\boldsymbol{\partial_{\phi}}(H_{\text{free}}+H_{\text{grad}})dt+d\boldsymbol{\boldsymbol{\pi}}_{\text{scatt}}+d\boldsymbol{\boldsymbol{\pi}}_{\text{coll}},\label{eq:dpi}
\end{alignat}
where $H_{\text{free}}$ and $H_{\text{grad}}$ denote the free
Hamiltonian and the gradient potential, respectively,
$d\boldsymbol{p}_{\text{scatt}}$,
$d\boldsymbol{\boldsymbol{\pi}}_{\text{scatt}}$ denote the
non-conservative terms induced by photon scattering, and
$d\boldsymbol{p}_{\text{coll}}$,
$d\boldsymbol{\boldsymbol{\pi}}_{\text{coll}}$ the non-conservative
terms, which arise from gas collisions. Specifically,
$d\boldsymbol{p}_{\text{scatt}}$ corresponds to the radiation pressure
scattering force, which displaces the particle along the positive $z$
direction, and $d\boldsymbol{p}_{\text{coll}}$ denotes the terms,
which tend to thermalize the center of mass motion to the temperature
$T$ of the gas of particles. Similarly,
$d\boldsymbol{\boldsymbol{\pi}}_{\text{scatt}}$ denotes the terms,
which quantify the transfer of angular momentum from the photons to
the particle, i.e. the driving terms, and
$d\boldsymbol{\pi}_{\text{coll}}$ denotes the terms that tend to
thermalize rotations to the temperature of the gas, i.e. the friction
and diffusive terms (see \cite{supplement}).

We consider a specific experimental situation, where we illustrate the
physical content of Eqs.~(\ref{eq:dr})-(\ref{eq:dpi}), and we obtain
approximate expression for the dominant mechanical frequencies
(further details can be found elsewhere
\cite{supplement}). Specifically, we consider the experimental
situation that produces the power spectral density in
Fig.~\ref{fig:psd}, where the rotational frequencies are significantly
higher or lower than the translational ones. To obtain the dominant
mechanical frequencies, we can in first approximation treat
translation and rotation as decoupled motion.

We start by looking at translational degrees of freedom. We suppose
that $\vert\frac{\mathbf{r}}{\lambda}\vert\ll1$, where $\lambda$
is the laser wavelength, which limits translations to harmonic oscillations.
In particular, the frequencies for the $x$, $y$, $z$ motion are
given by:
\begin{alignat}{2}
\omega_{x}^{2}=\frac{2Pa_{1}\chi_{0}}{c\sigma_{L}w_{0}^{2}\rho}, & \quad\omega_{y}^{2}=\frac{2Pa_{2}\chi_{0}}{c\sigma_{L}w_{0}^{2}\rho},\quad & \omega_{z}^{2}=\frac{2P\chi_{0}}{c\sigma_{L}\rho z_{R}^{2}},\label{eq:translations}
\end{alignat}
respectively, where $P$ is the laser power, $\sigma_{L}=\pi w_0^2$  is the
effective laser beam cross section area, $w_0$ is the mean beam waist radius, $a_{1}$,$a_{2}$ quantify the
asymmetry of the beam along the $x$, $y$ directions, respectively,
$z_{R}$ is the Rayleigh length, $\rho$ is the particle density,
$\chi_{0} = \frac{1}{3}\sum_{i=1}^3 \chi_i$ is an effective susceptibility
of the particle , and $c$ is the speed of light. These frequencies are
obtained directly from $H_{\text{grad}}$ by expanding to order
$\mathcal{O}((\frac{\mathbf{r}}{\lambda})^{2})$.

The rotational frequencies arise from (i) the transfer of angular
momentum during photon scattering, and from (ii) the gradient torque.
On the one hand, the scattering torque drives the system into a fast
spinning motion, while, on the other hand, the gradient torque would
like to align the system with the polarization of the incoming beam in
such a way to minimize the electric dipole potential energy, resulting
in nutation and precession. Before deriving the rotational frequencies
mathematically we now first give an intuitive picture of the two
mechanisms.

The mechanism (i) can be understood in terms of the angular momentum
carried by the incoming light beam (in a particle picture one can
think of an individual photon carrying a small amount of angular
momentum, e.g. $\hbar$ for circular polarization). During scattering
the angular momentum is transferred to the nanoparticle, where the
amount that is transferred depends on the susceptibility anisotropy
and orientation of the nanoparticle. As a consequence, the particle
starts to spin, until an asymptotic rotational frequency is reached,
which is constrained by friction due to gas collisions. In particular,
we consider the experimental situation, where the photon scattering
gives rise to high spinning frequencies
$\omega_\alpha^{\text{(spin)}}$ and $\omega_\gamma^{\text{(spin)}}$
about the $z$ and $z''$ axis, respectively.

Besides the dominant spinning motions there are also two additional
secondary motions, which arise as a consequence of the mechanism
(ii). In a nutshell, the gradient torque would like to align the
nanoparticle in such a way to minimize the electric dipole potential
energy, i.e. $\beta_0=\frac{\pi}{2}$, but once the nanoparticle starts
to spin, i.e. it acquires a large angular momenta along the $z$ and
$z''$ axis, it is unable to fully align, but rather settles around an
equilibrium position $\beta_0\neq \frac{\pi}{2}$, which can be readily
understood in terms of angular momenta addition. Any small
perturbation, e.g. gas collisions, will make the $\beta$ angle
oscillate around the $\beta_0$ angle, which results in libration
(nutation) motion with frequency
$\omega_\beta^{\text{(nutation)}}$. In addition, the coupling between
$\beta$ and $\alpha$ also creates a second frequency for the $\alpha$
motion, which we denote by $\omega_\alpha^{\text{(precession)}}$: this
motion can be visualized as a slow precession of the $z''$ axis about
the $z$ axis (see Fig~\ref{fig:theory} (b)). We now derive the
rotational frequencies.

The spinning frequencies can be obtained from Eq.~\eqref{eq:dpi}, by
setting $d\boldsymbol{\boldsymbol{\pi}}=0$, $\beta=\beta_{0}$, and
neglecting conservative and stochastic terms, i.e. we consider only
the driving term due to photon scattering and the friction term due to
gas collisions. We find the asymptotic angle conjugate momentum
$\boldsymbol{\boldsymbol{\pi}}^{\text{(spin)}}=-\frac{1}{2\Gamma_{c}}\boldsymbol{N}_{s}$,
where $\Gamma_{c}$ is the collisional damping rate, and
$\boldsymbol{N}_{s}=({N}_{\alpha},{N}_{\beta},{N}_{\gamma})^\top$ is
the photon scattering torque. We then immediately find the spinning
frequencies:
\begin{alignat}{1}
(\omega_{\alpha}^{\text{(spin)}},0,\omega_{\gamma}^{\text{(spin)}})^{\top} & =\mathbb{E}[Y]\boldsymbol{\pi}^{\text{(spin)}},\label{eq:spin}
\end{alignat}
where $\mathbb{E}[\,\cdot\,]$ denotes the time-average over fast
oscillating terms, $Y=Y(I)$ is the matrix that maps $\boldsymbol{\pi}$
to $\dot{\boldsymbol{\phi}}$, and $I$ is the moment of inertia tensor
in the body-frame (see \cite{supplement}). The explicit expressions
for $\omega_{\alpha}^{\text{(spin)}}$ and
$\omega_{\gamma}^{\text{(spin)}}$ are given in Eqs.~(B3) and (B5),
respectively.

\begin{figure*}[t]
\centering \includegraphics[clip,width=1\textwidth]{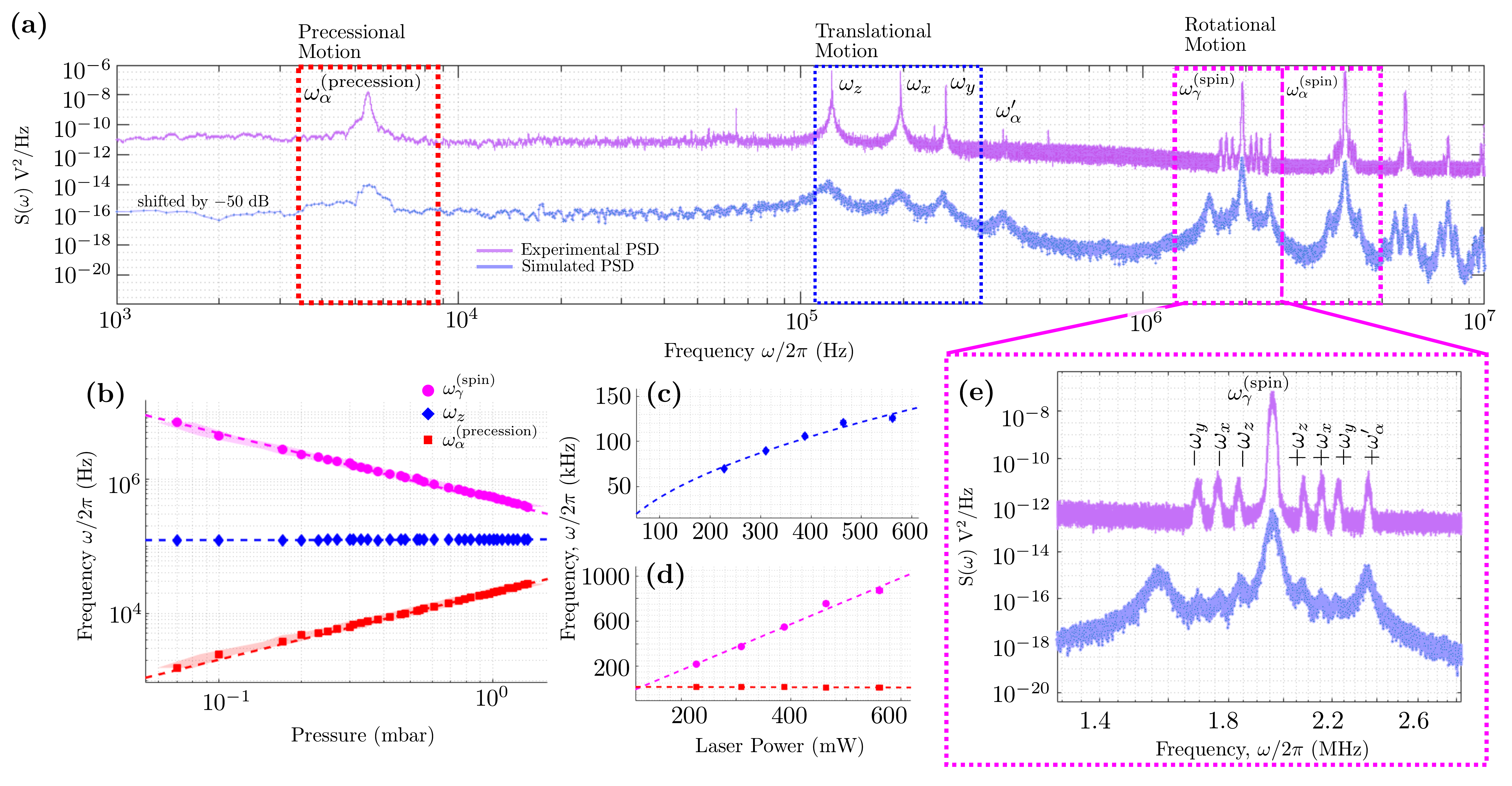}
\caption{\textbf{Measured Spectrum:} \textbf{(a)} The full
  experimental spectrum obtained at $1 \times 10^{-1}$ mbar (shown in
  purple).  It includes contributions from translation, rotation and
  precession motion. Additionally, a simulated spectrum (shown in
  blue) fitted to the experimental PSD is shifted for visibility. The
  simulated spectrum is obtained for $10^{8}$ timesteps. \textbf{(b)}
  Shows the frequency dependencies with pressure for translation
  (blue), rotation (magenta) and precession (red). The inverse
  relation is signature of rotational, while the direct
  proportionality is signature of precession.  \textbf{(c)} Shows the
  change in translation frequency due to square root of laser
  power. \textbf{(d)} shows linear dependency of rotation,
  $\omega_{\gamma}^{\text{(spin)}}$, with laser power whilst
  precession is independent of laser power. \textbf{(e)} Detailed
  spectrum of rotation frequency, $\omega_{\gamma}^{\text{(spin)}}$,
  with translation sidebands and $\omega_{\alpha}^{'}$ an additional
  frequency peak appearing in the motion of $\alpha$ motion.}
\label{fig:psd} 
\end{figure*}

We now consider the oscillations of $\beta$ about the equilibrium
point $\beta_{0}$, which we denote by $\delta\beta_{0}$. This
oscillatory, libration motion is induced by the (ii) conservative
terms, as well as by the fast spinning motion of $\alpha$ and
$\gamma$. In particular, after writing the Hamiltonian eqs. of motion
for $\delta\beta_{0}$, using Eqs.~\eqref{eq:dphi} and \eqref{eq:dpi},
performing the time average $\mathbb{E}[\,\cdot\,]$, and keeping only
the dominant terms, we eventually obtain:
\begin{equation}
  \omega_{\beta}^{\text{nutation}}=\frac{1}{2}\frac{I_{1}+I_{2}}{I_{1}I_{2}}\csc ^2(\beta_0)\pi_{\alpha}^{\text{(spin)}},\label{eq:nutation}
\end{equation}
where $I_{1}$ and $I_{2}$ denote the moment of inertia along the $x''$
and $y''$ principal axis. The explicit expressions for
$\omega_{\beta}^{\text{(nutation)}}$ and $\beta_0$ are given in
Eqs.~(B5) and (B6), respectively.

The $\delta\beta_0$ oscillations also perturb the $\alpha$ motion: we
denote the perturbation to the spinning motion by $\delta\alpha$,
i.e. $\alpha(t)=\omega_\alpha^{\text{(spin)}} t+\delta\alpha$. In
particular, from Eq.~\eqref{eq:dphi}, performing time-average
$\mathbb{E}[\,\cdot\,]$, and using Eq.~\eqref{eq:nutation}, we
eventually obtain
$
\dot{\delta\alpha}=2\omega_\beta^{\text{(nutation)}}\text{cot}(\beta_0)\delta
\beta$, where
$\delta
\beta=\mathcal{B}\text{cos}(\omega_\beta^{\text{(nutation)}}t)$, and
$\mathcal{B}$ denotes the amplitude of $\delta \beta$
oscillations. However, in a typical detection we do not measure
directly $\alpha$, but normally its sine or cosine value. We consider
here $\text{sin}(\alpha)$, which we Taylor expand to order
$\mathcal{O}(\delta\alpha)$, i.e.
$\text{sin}(\alpha)=\text{sin}(\omega_\alpha^{\text{(spin)}}
t)+\text{cos}(\omega_\alpha^{\text{(spin)}} t)\delta\alpha$. In
particular, from the last term, using trigonometric identities, and
the expressions for $\delta\alpha$ and $\delta\beta_0$, we obtain a
term proportional to
$\text{sin}(\omega_{\alpha}^{\text{(precession)}}t)$, where
\begin{equation}
  \omega_{\alpha}^{\text{(precession)}}=\omega_\alpha^{\text{(spin)}}-\omega_\beta^{\text{(nutation)}}. \label{eq:precession}
\end{equation}
The $\alpha$ degree of freedom has thus two distinct motions: a fast
spinning motion with the frequency given in Eq.~(\ref{eq:spin}) and a
slow precession motion with the frequency given in
Eq.~(\ref{eq:precession}). This precession motion can be seen as a
consequence of the $\beta$ motion, which perturbs back the $\alpha$
motion.

From Eqs.~(\ref{eq:spin}) and (\ref{eq:nutation}), noting that
$\boldsymbol{\pi}^\text{(spin)} \propto
\frac{\boldsymbol{N}_{s}}{\Gamma_c} \propto \frac{P}{p}$, we find that
$\omega_\alpha^{\text{(spin)}}$ and $\omega_\beta^{\text{(nutation)}}$
scale linearly with the laser power $P$, and are inversely
proportional to the gas pressure $p$, where we have assumed that the
equilibrium position $\beta_0$ does not change significantly near an
initially chosen power $P_0$ and pressure $p_0$, $\boldsymbol{N}_{s}$
is the torque due to photon scattering, and $\Gamma_c$ is the damping
rate due to gas collisions. On the other hand, the precession
frequency $\omega_{\alpha}^{\text{(precession)}}$ given in
Eq.~(\ref{eq:precession}) can scale differently depending on the
values of $\partial_P \Delta \omega\vert_{P_0}$ and
$\partial_p \Delta \omega\vert_{p_0}$, where
$\Delta
\omega=\omega_\alpha^{\text{(spin)}}-\omega_\beta^{\text{(nutation)}}$.

We have confirmed the validity of the obtained approximate formulae in
Eqs.~(\ref{eq:translations})-(\ref{eq:precession}) by numerically
simulating Eqs.~(\ref{eq:dr})-(\ref{eq:dpi}). More details of the
simulations will be discussed elsewhere \cite{Toros2018s}.

{\it Experiments--} The optical trap is shown in
Fig. \ref{fig:theory}(a). In the experiments presented we use
initially individual silica nanoparticles dispersed in water. We
observe ageing of the solution with the result of clustering of the
nanospheres into compounds of two to three nanospheres, a few weeks
after preparation, as shown in Fig. \ref{fig:theory}(c). The aged
solution is delivered to the trap using a nebuliser. The optical
scattering force, limits the maximum particle mean radius that can be
optically trapped to about 150 nm to 200 nm.

{\it Results--} The power spectral density (PSD) shown in
Fig. \ref{fig:psd}(a) is generated from the time trace recorded as the
photodetector signal, $I_{\text{exp}}$, over a time interval of one
second. The PSD shows a rich spectrum of frequencies, which respond
differently for changing laser power and background gas pressure.

Translation motion is observed at frequencies, $\omega_{x}$ =
$2\pi \times$ 196 kHz, $\omega_{y}$ = $2\pi \times$ 246 kHz,
$\omega_{z}$ = $2\pi \times$ 124 kHz in Fig.~\ref{fig:psd}(a). We do
not observe pressure dependency of the translation frequencies (see
Fig.~\ref{fig:psd}(b)), but find that they scale proportionately to
the square root of laser power, $P$, in agreement with
Eq.(\ref{eq:translations}), see Fig. \ref{fig:psd}(c). The $x$ and $y$
peaks are separated because we use elliptically polarized light, which
after it is reflected from the paraboloidal mirror, generates an
asymmetric optical trap. The polarization of light was kept constant
during the course of the experiments in this letter.

From the experimental data, we find the fundamental frequencies for
the rotational motions, $\omega_{\gamma}^{\text{spin}}$ and
$\omega_{\alpha}^{\text{spin}}$ to be $2\pi \times 1.9$ MHz and
$2\pi \times 3.8$ MHz, respectively, which are perfectly reproduced
using Eqs. (\ref{eq:spin}), (B3) and (B5). The rotational frequency
$\omega_{\gamma}^{(\text{spin})}$ changes with the damping,
$\Gamma_c$, which depends linearly on the gas pressure, $p$, i.e.
$\omega_{\gamma}^{\text{(spin)}} \propto \frac{1}{\Gamma_c} \propto
\frac{1}{p} $. This is a clear signature of rotation motion, as shown
in Fig. \ref{fig:psd}(b). We also observe a dependency of the
frequency on the laser power $P$, as shown in Fig \ref{fig:psd}(d) in
agreement with the dependency on the photon scattering torque,
$\boldsymbol{N}_s \propto P$. Zooming to the fundamental frequency
$\omega_{\gamma}^{(\text{spin})}$ reveals sidebands, see
Fig.~\ref{fig:psd}(e), which are the addition and subtraction of the
three translational frequencies. Using the numerical simulation we
identify another mode in $\alpha$ rotation, with frequency
$\omega_{\alpha}^{'} = 2 \pi \times 393$ kHz: this gives rise to the
sideband in $\omega_{\gamma}^{(\text{spin})}$ (see
\cite{supplement}). In the presented data set, we only resolve one of
the $\omega_{\alpha}^{'}$ sideband peaks. The light-matter
interactions introduce couplings between translation and rotation,
which explain the observed sidebands and higher harmonics in agreement
with numerical simulation, see Eqs.~(A2), (A5)-(A7) in
\cite{supplement} for further details. We also observe
$\omega_{\alpha}^{(\text{spin})}$ to scale linearly with power and
inversely with pressure.

Further to observing translation and rotation peaks, we observe a
frequency at $2\pi \times 5.4$ kHz, as shown in
Fig.~\ref{fig:psd}(a). This frequency is well-isolated and
characterised by its dependency on laser power, $P$ and gas pressure,
$p$. We associate this low frequency with the gradient torque-induced
precession, $\omega_{\alpha}^{\text{(precession)}}$. From
Eq.~(\ref{eq:precession}), Taylor expanding about the initial pressure
$p_0$, we find that the dominant term is linear in $p$ (the constant
terms cancel), i.e.
$\omega_{\alpha}^{\text{(precession)}} =
\partial_p(\Delta\omega)\vert_{p_0} p$. On the other hand, we find a
weak dependence on the laser power $P$,
i.e. $\omega_{\alpha}^{\text{(precession)}} =\Delta\omega\vert_{P_0}$,
where $P_0$ is the initial laser power. We verify these frequency
dependencies on gas pressure and laser power in Figs.~\ref{fig:psd}(b)
and \ref{fig:psd}(d), respectively. We exclude
$\omega_{\alpha}^{\text{(precession)}}$ to be caused by nonlinear
translational motion, as the translation frequencies do not change
with pressure. This is evident from Fig.~\ref{fig:psd}(b) where
$\omega_{z}$ is constant. Thus, we conclude, that the observed
frequency and its behaviour is signature of precession motion as
described by Eq.~(\ref{eq:precession}).

The precession motion arises due to the fast spinning $\alpha$ degree
of freedom, as well as, the nutation motion of $\beta$. The
observation of precession is thus an indirect indication of nutation
motion. We associate the time trace of the $\beta$ motion to libration
motion, which is linearly dependant on power and inverse proportional
to gas pressure, as a result of the numerical simulation. The $\beta$
libration is due to the coupling between $\alpha$ and $\beta$ motion.

{\it Discussion--} From the theoretical analysis, the precession
motion arises due to an optical-torque acting upon the trapped
particle. Torque can also be generated by an external force, which
opens the way for sensitive detection of forces by precession.  In
particular, by combining Eqs.~(\ref{eq:spin}), (\ref{eq:nutation}),
and (\ref{eq:precession}), we get an expression for the
$\alpha$-component of the photon scattering torque:
\begin{equation}
  N_\alpha=  \Gamma_c \sin^2(\beta_0) \left[\frac{2 I_1 I_2 (I_1 + I_2)}{I_1^2 + I_2^2} \right]
  \omega_{\alpha}^{\text{(precession)}},
   \label{nalpha}
 \end{equation}
 where we have kept only the dominant terms, and denote the term in
 the square brackets by $\mathcal{J}$, and name it the effective
 moment of inertia. Using the experimentally measured
 $\omega_{\alpha}^{\text{(precession)}}$, and estimating
 $\mathcal{J}$, we achieve a measured torque of
 $N_\alpha= 1.9 \pm 0.5 \times 10^{-23}\text{Nm}$ at
 $1 \times 10^{-1}$ mbar, in comparison to the measurement of
 nanoscale torque sensors reported down to $10^{-20}$ Nm
 \cite{kim2013nanoscale} and to estimates of $10^{-22}$ Nm
 \cite{Kuhn2017} for silicon nanorods.

 We now consider an additional small external torque acting on
 $\alpha$, which we denote by $\delta N_{\text{ext}}$.  We suppose
 that Eq.~\eqref{nalpha} remains valid, when we formally make the
 replacement $N_\alpha\rightarrow N_\alpha+\delta N_{\text{ext}}$, and
 we denote the corresponding change in the precession frequency by
 $\delta \omega_{\alpha}$. Furthermore, assuming that the equilibrium
 value of $\beta_0$ remains largely unaffected, we obtain the torque
 sensitivity
 $\delta N_{\text{ext}}=\Gamma_c\text{sin}(\beta_0)^2\mathcal{J}\delta
 \omega_{\alpha}$. Using experimental parameters we estimate a torque
 sensitivity of $3.6 \pm 1.1 \times 10^{-31}$ Nm$/\sqrt{\text{Hz}}$ at
 $1 \times 10^{-7}$ mbar, limited only by photon shot noise (see
 supplementary D \cite{supplement}) in comparison to the torque
 sensitivity using libration motion, which is at $1 \times 10^{-29}$
 Nm$/\sqrt{\text{Hz}}$ at $10^{-9}$ mbar \cite{Hoang2016}.

 {\it Conclusions--} We have observed precession motion in levitated
 optomechanics and inferred the presence of nutation motion. We
 present a theory of this motion and show that it arises from the
 equations of motion for a rotating object experiencing scattering and
 gradient forces and torques. Additionally, we characterise the rich
 spectrum detected with translation, rotation and higher harmonics. We
 further show a measured torque of $ 10^{-23}$ Nm at
 $10^{-1}$ mbar, and predict the ability to reach
 sensitivities down to $10^{-31}$ Nm$/\sqrt{\text{Hz}}$ at
 $10^{-7}$ mbar.

 Thus, precession motion is a degree of freedom that could be utilised
 for torque sensing with the sensitivities to resolve single electron
 \cite{Rugar2004} and even nuclear spins \cite{Hoang2016} at low
 pressure. This work paves the way for gyroscope applications, as
 shown in \cite{nagornykh2017optical}. The precession motion can also
 be used for dynamical model selection to distinguish between quantum
 and classical evolution due to the inherent nonlinearities in
 rotation motion \cite{ralph2017dynamical}, if sufficient coherence
 can be prepared. Additionally, this work can be used to reconstruct
 the shape and effective moment of inertia, $\mathcal{J}$, from the
 knowledge of the full spectrum of freedom containing all motional
 degrees of the trapped particle.

 \textit{Acknowledgments\textendash{}} We thank C. Timberlake and
 G. Winstone for discussion and M. Rademacher for assistance with the
 TEM image.  We thank for funding the Leverhulme Trust
 {[}RPG-2016-046{]} and the EU Horizon 2020 research and innovation
 programme under grant agreement No 766900 [TEQ].  A.S. is supported
 by the UK Engineering and Physical Sciences Research Council (EPSRC)
 under Centre for Doctoral Training Grant No. EP/L015382/1. All data
 supporting this study are openly available from the University of
 Southampton repository at https://doi.org/10.5258/SOTON/D0523"

\bibliographystyle{apsrev4-1}
\bibliography{rotation}

\appendix
\widetext
\newpage
\title{Supplementary Material for Precession Motion in Levitated Optomechanics}

\author{Muddassar Rashid}
\email{m.rashid@soton.ac.uk}

\selectlanguage{english}%

\affiliation{Department of Physics and Astronomy, University of Southampton, SO17
1BJ, UK}

\author{Marko Toro\v{s}}
\email{m.toros@soton.ac.uk}

\selectlanguage{english}%

\affiliation{Department of Physics and Astronomy, University of Southampton, SO17
1BJ, UK}

\author{Ashley Setter}

\affiliation{Department of Physics and Astronomy, University of Southampton, SO17
1BJ, UK}

\author{Hendrik Ulbricht}
\email{h.ulbricht@soton.ac.uk}

\selectlanguage{english}%

\affiliation{Department of Physics and Astronomy, University of Southampton, SO17
  1BJ, UK}

\maketitle

\clearpage \onecolumngrid 
\section{Dynamics }

In this supplementary section we list the terms obtained in~\cite{Toros2018t,Toros2018s}.
We start by specifying the conservative terms. The free Hamiltonian
is given by:

\begin{alignat}{1}
H_{\text{free}}=\frac{p_{x}{}^{2}+p_{y}{}^{2}+p_{z}{}^{2}}{2M}+\bigg( & \frac{\csc^{2}(\beta)(\cos(\gamma)(\pi_{\alpha}-\pi_{\gamma}\cos(\beta))-\text{\ensuremath{\pi_{\beta}}}\sin(\beta)\sin(\gamma))^{2}}{2I_{1}}\nonumber \\
 & +\frac{\csc^{2}(\beta)(\sin(\gamma)(\text{\ensuremath{\pi_{\alpha}}}-\pi_{\gamma}\cos(\beta))+\pi_{\beta}\sin(\beta)\cos(\gamma))^{2}}{2I_{2}}+\frac{\pi_{\gamma}{}^{2}}{2I_{3}}\bigg),\label{eq:Hfree}
\end{alignat}
where $M$ is the mass of the nanoparticle, and $I=\text{diag}(I_{1},I_{2},I_{3})$
is the moment of inertia tensor in the body frame. 

The gradient potential is given by

\begin{alignat}{1}
H_{\text{grad}}=-\frac{VP}{c\sigma_{L}}\vert u(\boldsymbol{r})\vert^{2} & \bigg(a^{2}\big(\text{\ensuremath{\chi_{1}}}(\cos(\alpha)\cos(\beta)\cos(\gamma)-\sin(\alpha)\sin(\gamma))^{2}\nonumber \\
 & \qquad\qquad+\text{\ensuremath{\chi}}_{2}(\cos(\alpha)\cos(\beta)\sin(\gamma)+\sin(\alpha)\cos(\gamma))^{2}+\text{\ensuremath{\chi}}_{3}\cos^{2}(\alpha)\sin^{2}(\beta)\big)\nonumber \\
 & \quad+b^{2}\bigg(\text{\ensuremath{\chi}}_{1}(\sin(\alpha)\cos(\beta)\cos(\gamma)+\cos(\alpha)\sin(\gamma))^{2}\nonumber \\
 & \qquad\qquad+\text{\ensuremath{\chi}}_{2}(\cos(\alpha)\cos(\gamma)-\sin(\alpha)\cos(\beta)\sin(\gamma))^{2}+\text{\ensuremath{\chi}}_{3}\sin^{2}(\alpha)\sin^{2}(\beta)\big)\bigg).\label{eq:Hgradient}
\end{alignat}
where $u$ is a modified Gaussian mode function:

\begin{equation}
u(\boldsymbol{r})=\frac{w_{0}}{w(z)}\text{exp}\left(-\frac{a_{1}x^{2}+a_{2}y^{2}}{w(z)^{2}}\right)e^{-ikz},\label{eq:umode}
\end{equation}
$w_{0}$ is an effective beam waist, $\sigma_{L}=\pi w_{0}^{2}$,
$a_{1}$, $a_{2}$ denote the asymmetry along the $x$, $y$ axis,
respectively, $a_{1}a_{2}=1$, $w(z)=w_{0}\sqrt{(1+\frac{z^{2}}{z_{R}^2})}$,
$z_{R}$ is the Rayleigh range, $k=\frac{2\pi}{\lambda}$, $\lambda$
is the laser wavelength, $P$ is the laser power, $c$ is the speed
of light, and $\chi=\text{diag}(\chi_{1},\chi_{2},\chi_{3})$ is the
susceptibility tensor in the body-frame. 

We now specify the non-conservative terms. We first discuss the deterministic
terms related to photon scattering, namely $d\boldsymbol{\boldsymbol{p}}_{\text{scatt}}=(0,0,dp_{z}^{\text{(ds)}})^\top$
and $d\boldsymbol{\boldsymbol{\pi}}_{\text{scatt}}=(d\pi_{\alpha}^{\text{(ds)}},d\pi_{\beta}^{\text{(ds)}},d\pi_{\gamma}^{\text{(ds)}})^\top$. In particular, we have:

\begin{alignat}{1}
dp_{z}^{\text{(ds)}} & =\frac{16\pi\hbar\Gamma_{s}}{3}\frac{2\pi}{\lambda}\vert u(\boldsymbol{r})\vert^{2}dt,\label{eq:sforce}\\
d\pi_{\alpha}^{\text{(ds)}} & =\frac{4\pi b_{x}b_{y}\hbar\Gamma_{s}}{3}\vert u(\boldsymbol{r})\vert^{2}\bigg(-2\sin^{2}(\beta)\cos(2\gamma)(\text{\ensuremath{\chi_{1}}}-\text{\ensuremath{\chi}}_{2})(\text{\ensuremath{\chi}}_{1}+\text{\ensuremath{\chi}}_{2}-2\text{\ensuremath{\chi}}_{3})\nonumber \\
 & \qquad\qquad\qquad\qquad\qquad\qquad\,\,\,\,\,+\cos(2\beta)\left(\text{\ensuremath{\chi}}_{1}^{2}+2\text{\ensuremath{\chi}}_{3}(\text{\ensuremath{\chi}}_{1}+\text{\ensuremath{\chi}}_{2})-4\text{\ensuremath{\chi}}_{1}\text{\ensuremath{\chi}}_{2}+\text{\ensuremath{\chi}}_{2}^{2}-2\text{\ensuremath{\chi}}_{3}^{2}\right)\nonumber \\
 & \qquad\qquad\qquad\qquad\qquad\qquad\,\,\,\,\,+3\text{\ensuremath{\chi}}_{1}^{2}-2\text{\ensuremath{\chi}}_{3}(\text{\ensuremath{\chi}}_{1}+\text{\ensuremath{\chi}}_{2})-4\text{\ensuremath{\chi}}_{1}\text{\ensuremath{\chi}}_{2}+3\text{\ensuremath{\chi}}_{2}^{2}+2\text{\ensuremath{\chi}}_{3}^{2}\bigg)dt,\label{eq:ds1}\\
d\pi_{\beta}^{\text{(ds)}} & =\frac{16\pi b_{x}b_{y}\hbar\Gamma_{s}}{3}\vert u(\boldsymbol{r})\vert^{2}\sin(\beta)\sin(\gamma)\cos(\gamma)(\text{\ensuremath{\chi}}_{1}-\text{\ensuremath{\chi}}_{2})(\text{\ensuremath{\chi}}_{1}+\text{\ensuremath{\chi}}_{2}-2\text{\ensuremath{\chi}}_{3})dt,\label{eq:ds2}\\
d\pi_{\gamma}^{\text{(ds)}} & =\frac{16\pi b_{x}b_{y}\hbar\Gamma_{s}}{3}\vert u(\boldsymbol{r})\vert^{2}\cos(\beta)(\text{\ensuremath{\chi}}_{1}-\text{\ensuremath{\chi}}_{2})^{2}dt,\label{eq:ds3}
\end{alignat}
where $\Gamma_{s}=\frac{\tilde{\sigma}_{R}}{\sigma_{L}}\frac{P}{\hbar\omega_{L}}$
is the scattering rate, $\omega_{L}=\frac{2\pi c}{\lambda}$, $\tilde{\sigma}_{R}=\frac{\pi^{2}V_{0}^{2}}{\lambda^{4}}$
is an effective scattering cross-section, $b_{x}$ and $b_{y}$ are
the components of the Gaussian beam polarization vector $\boldsymbol{\boldsymbol{\epsilon}}_{d}=(b_{x},ib_{y},0)^{\top}$,
and $b_{x}^{2}+b_{y}^{2}=1$. 

We now discuss the gas collision terms denoted by $d\boldsymbol{\boldsymbol{p}}_{\text{coll}}=d\boldsymbol{p}^{\text{(dc)}}+d\boldsymbol{p}^{\text{(sc)}}$
and $d\boldsymbol{\boldsymbol{\pi}}_{\text{coll}}=d\boldsymbol{\boldsymbol{\pi}}^{\text{(dc)}}+d\boldsymbol{\boldsymbol{\pi}}^{\text{(sc)}}$. The terms denoted by the superscripts (ds) and (sc) corresponds to non-conservative deterministic and stochastic terms, respectively. In particular, we have
\begin{alignat}{1}
  d\boldsymbol{p}^{\text{(dc)}} & =-2\Gamma_{C}\boldsymbol{p}dt,\label{eq:Pd}\\
  d\boldsymbol{\boldsymbol{\pi}}^{\text{(dc)}} & =-2\Gamma_{C}\boldsymbol{\boldsymbol{\pi}}dt,\label{eq:Pid}\\
  d\boldsymbol{p}_{k}^{\text{(sc)}} & =\sqrt{4mk_{B}T\Gamma_{c}}dV_{k},\label{eq:Pnc}\\
  d\boldsymbol{\boldsymbol{\pi}}_{k}^{\text{(sc)}} &
  =\sum_{\zeta,j=1}^{3}\sqrt{4k_{B}mT\tilde{D}_{\zeta}\Gamma_{c}}(\partial_{\mathbf{\mathbf{\phi}}_{k}}R)_{j,\zeta}dZ_{\zeta,j},\label{eq:Pinc}
\end{alignat}
where $\Gamma_{c}=\frac{\pi p_{g}r_{g}^{2}}{\sqrt{8m_{g}k_{B}T}}$
is a characteristic collision rate, $p_{g}$ is the gas pressure,
$r_{g}$ and $m_{g}$ are the radius and mass of a gas particle, respectively,
$k_{B}$ is Boltzmann's constant, $T$ is the gas temperature, and
$\tilde{D}_{\zeta}=\frac{1}{2}(\text{tr}(I)-I_{\zeta})$. $V_{k}$
and $Z_{\zeta,j}$ are zero mean independent Wiener processes.

\section{System frequencies}

Here we list the formulae, which have been used in the main text, to obtain the dominant frequencies of the system. The photon scattering
torque is given by

\begin{equation}
\boldsymbol{N}_{s}=\mathbb{E}\left[\left(\frac{d\pi_{\alpha}^{\text{(ds)}}}{dt},\frac{d\pi_{\beta}^{\text{(ds)}}}{dt},\frac{d\pi_{\gamma}^{\text{(ds)}}}{dt}\right)^{\top}\right],
\end{equation}
where $d\pi_{\alpha}^{\text{(ds)}}$,$d\pi_{\beta}^{\text{(ds)}}$,
and $d\pi_{\gamma}^{\text{(ds)}}$ are given in Eqs.~(\ref{eq:ds1}),
(\ref{eq:ds2}), and (\ref{eq:ds3}), respectively, and $\mathbb{E}$
denotes the time average over fast oscillating terms.

The conversion matrix $Y$ from the conjugate angle momenta $\boldsymbol{\boldsymbol{\pi}} $ to the time-derivative
of the angle vector $\dot{\boldsymbol{\boldsymbol{\phi}}}$ is defined as
\begin{equation}
Y=\left(\left(\begin{array}{ccc}
-\sin(\beta)\cos(\gamma) & \sin(\gamma) & 0\\
\sin(\beta)\sin(\gamma) & \cos(\gamma) & 0\\
\cos(\beta) & 0 & 1
\end{array}\right)^{\top}\left(\begin{array}{ccc}
I_{1} & 0 & 0\\
0 & I_{2} & 0\\
0 & 0 & I_{3}
\end{array}\right)\left(\begin{array}{ccc}
-\sin(\beta)\cos(\gamma) & \sin(\gamma) & 0\\
\sin(\beta)\sin(\gamma) & \cos(\gamma) & 0\\
\cos(\beta) & 0 & 1
\end{array}\right)\right)^{-1}.
\end{equation}
The dominant frequencies for the spin motion are given by
\begin{alignat}{1}
\omega_{\alpha}^{\text{(spin)}}= & \frac{2b_{x}b_{y}\pi\hbar}{\frac{3}{4}\left(I_{1}^{2}+2I_{1}I_{2}+I_{2}^{2}\right)I_{3}}\frac{\text{\ensuremath{\Gamma_{s}}}}{\Gamma_{c}}\csc^{2}(\text{\ensuremath{\beta_{0}}})\bigg[\frac{1}{2}(I_{1}+I_{2})I_{1}(\text{\ensuremath{\chi_{1}}}-\text{\ensuremath{\chi_{2}}})^{2}\cos^{2}(\text{\ensuremath{\beta_{0}}})-(\frac{1}{2}I_{1}I_{3}+\frac{1}{2}I_{2}I_{3})\nonumber \\
 & \left(3\text{\ensuremath{\chi_{1}^{2}}}-4\text{\ensuremath{\chi_{2}}}\text{\ensuremath{\chi_{1}}}+3\text{\ensuremath{\chi_{2}^{2}}}+2\text{\ensuremath{\chi_{3}^{2}}}-2(\text{\ensuremath{\chi_{1}}}+\text{\ensuremath{\chi_{2}}})\text{\ensuremath{\chi_{3}}}+\left(\text{\ensuremath{\chi_{1}^{2}}}-4\text{\ensuremath{\chi_{2}}}\text{\ensuremath{\chi_{1}}}+\text{\ensuremath{\chi_{2}^{2}}}-2\text{\ensuremath{\chi_{3}^{2}}}+2(\text{\ensuremath{\chi_{1}}}+\text{\ensuremath{\chi_{2}}})\text{\ensuremath{\chi_{3}}}\right)\cos(2\text{\ensuremath{\beta_{0}}})\right)\bigg], \label{alphaspin}
\end{alignat}
and
\begin{alignat}{1}
\omega_{\gamma}^{\text{(spin)}}= & \frac{2b_{x}b_{y}\pi\hbar}{\frac{3}{4}\left(I_{1}^{2}+2I_{1}I_{2}+I_{2}^{2}\right)I_{3}}\frac{\text{\ensuremath{\Gamma_{s}}}}{\ensuremath{\Gamma}_{c}}\cot(\text{\ensuremath{\beta}}_{0})\csc(\text{\ensuremath{\beta_{0}}})\nonumber \\
 & \bigg[-\frac{1}{2}\left((I_{1}+I_{2})I_{3}\cos^{2}(\text{\ensuremath{\beta_{0}}})+\left(\frac{1}{2}I_{1}^{2}+I_{2}I_{1}+\frac{1}{2}I_{2}^{2}\right)\sin^{2}(\text{\ensuremath{\beta_{0}}})\right)(\text{\ensuremath{\chi_{1}}}-\text{\ensuremath{\chi_{2}}})^{2}\bigg(-(-\frac{1}{2}I_{1}-\frac{1}{2}I_{2})I_{3}\nonumber \\
 & \left(3\text{\ensuremath{\chi_{1}^{2}}}-4\text{\ensuremath{\chi_{2}}}\text{\ensuremath{\chi_{1}}}+3\text{\ensuremath{\chi_{2}^{2}}}+2\text{\ensuremath{\chi_{3}^{2}}}-2(\text{\ensuremath{\chi_{1}}}+\text{\ensuremath{\chi_{2}}})\text{\ensuremath{\chi_{3}}}+\left(\text{\ensuremath{\chi_{1}^{2}}}-4\text{\ensuremath{\chi_{2}}}\text{\ensuremath{\chi_{1}}}+\text{\ensuremath{\chi_{2}^{2}}}-2\text{\ensuremath{\chi_{3}^{2}}}+2(\text{\ensuremath{\chi_{1}}}+\text{\ensuremath{\chi_{2}}})\text{\ensuremath{\chi_{3}}}\right)\cos(2\text{\ensuremath{\beta_{0}}})\right)\bigg)\bigg],
\end{alignat}
while the nutation frequency is given by 
\begin{alignat}{1}
\omega_{\beta}^{\text{nutation}} = \frac{1}{2}\frac{I_{1}+I_{2}}{I_{1}I_{2}}\csc ^2(\beta_0)
\frac{2 \pi  b_x b_y \hbar }{3 }\frac{\Gamma_s}{\Gamma_c}  
\bigg[& \cos (2 \beta_0 )
   \big( \chi_1^2+2 \chi_3
   (\chi_1+\chi_2)-4
   \chi_1 \chi_2+
   \chi_2^2-2 \chi_3^2\big) \nonumber \\
   &+3 \chi_1^2-2 \chi_3 
   (\chi_1+\chi_2)-4 \chi_1
   \chi_2+3 \chi_2^2
   +2 \chi_3^2\bigg]. \label{betanutation}
\end{alignat}
The equilibrium position of the $\beta$ angle is given approximately by:
\begin{equation}
\beta_0=\sin ^{-1}\left( \sqrt[4]{
	\frac{(I_1+I_2)\pi c  w_0^2 \pi_\alpha^2}
   {I_1 I_2 P V  (2 \chi_3-\chi_1-\chi_2)}}\right).\label{beta0}
\end{equation}
The fact that $\beta_0$ depends on $\pi_\alpha$ is a consequence of the coupling in Eqs.~\eqref{eq:Hfree} and \eqref{eq:Hgradient}. 

To estimate the dominant frequencies using Eqs.~\eqref{alphaspin} -
\eqref{betanutation} we have numerically simulated the system and
extracted the fitted parameters (see supplementary material
\ref{sec:SimSpectrum}). In particular, we obtain from the
simulation the following values
$\omega_{\alpha}^{\text{(spin)}}=2 \pi \times 3.919\text{MHz}$,
$\omega_{\gamma}^{\text{(spin)}}= 2 \pi \times 1.957 \text{MHz}$,
$\omega_{\beta}^{\text{(nutation)}}=2 \pi \times 3.924 \text{MHz}$,
and
$\omega_{\alpha}^{\text{(precession)}}=2 \pi \times 5.5 \text{kHz}$ in
perfect agreement with experimental data. Using the simulation
parameters, we also find good agreement within order of magnitude with
the approximate expressions in Eqs.~\eqref{alphaspin} -
\eqref{betanutation}. Thus, Eqs.~\eqref{alphaspin} -
\eqref{betanutation}, can be used for estimating the initial
simulation parameters for the fitting algorithm.

.

\section{Numerical Simulation}
\label{sec:SimSpectrum}

\begin{figure}[t]
  \centering
  \includegraphics[width=1\linewidth]{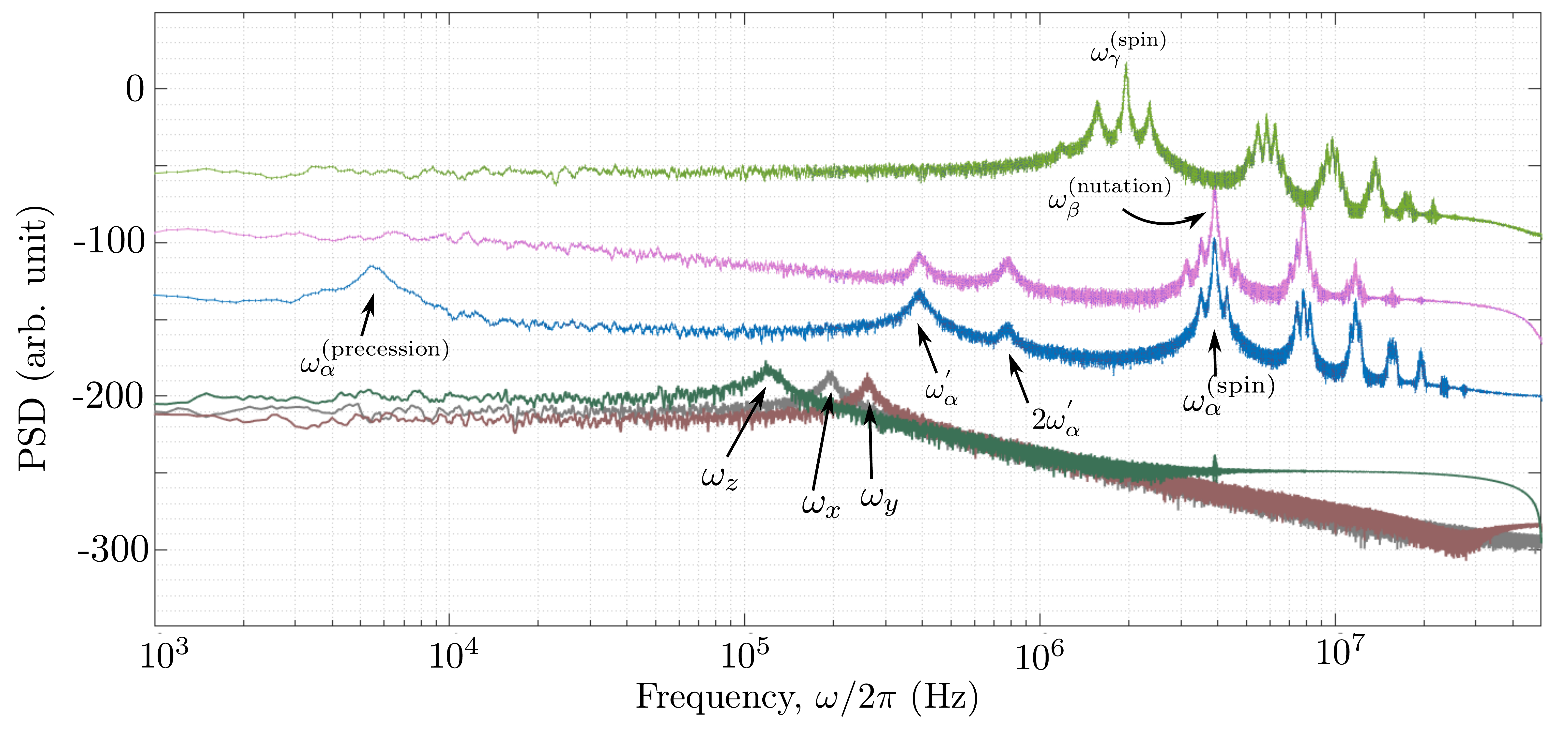}
  \caption{\label{fig:SimSpectrum} \textbf{Simulated Spectrum:} The
    power spectral density of the six different degrees of motion,
    where $\omega_{\gamma}^{\text{(spin)}}$ (in green),
    $\omega_{\beta}^{\text{(nutation)}}$ (in magenta),
    $\omega_{\alpha}^{\text{(spin)}}$ (in blue) are the rotation
    motions; $\omega_{x}$ (in grey), $\omega_{y}$ (in brown), and
    $\omega_{z}$ (in dark green) are the translation motions.}
\end{figure}

In this section we further discuss the different contributions to the
simulated spectrum shown in Fig.~2(a). The solution to the
twelve coupled SDEs, give information of $x(t)$, $y(t)$, $z(t)$,
$\alpha(t)$, $\beta(t)$ and $\gamma(t)$. From these we can extract the
frequency spectrum, as shown in Fig.~\ref{fig:SimSpectrum}. The
spectrum for each degree of freedom demonstrates the source of the
numerous frequencies observed in Fig.~2(a).

\emph{$\alpha$-rotation} (blue line in Fig.~\ref{fig:SimSpectrum}):
Starting from the right to left, the $\alpha$ rotation contains
frequencies relating the main rotation peak,
$\omega_{\alpha}^{\text{(spin)}}$ and its higher harmonics. In
addition to this there are sidebands to
$\omega_{\alpha}^{\text{(spin)}}$ which relate to another mode of
motion designated as $\omega_{\alpha}^{'}$. Farther to the left, of
the spectrum we observe this additional mode, $\omega_{\alpha}^{'}$
and its second harmonic. The far left of the $\alpha$ spectrum show
the precession motion.

\emph{$\beta$-nutation} (magenta line in Fig.~\ref{fig:SimSpectrum}):
The spectrum for $\beta$ motion shows the central nutation frequency
$\omega_{\beta}^{\text{(nutation)}}$ and its higher harmonics. The
sidebands refer to $\omega_{\alpha}^{'}$ and
$2\omega_{\alpha}^{'}$. Both these frequencies also appear in the
actual spectrum as well.

\emph{$\gamma$-rotation} (green line in Fig.~\ref{fig:SimSpectrum}):
The spectrum for $\gamma$ rotation shows
$\omega_{\alpha}^{\text{(spin)}}$ and its higher harmonics. The
sidebands relate to $\omega_{\alpha}^{'}$ and come about due to the
coupling of $\gamma$-rotation with $\alpha$-rotation. In addition to
the rotation degrees of freedom, Fig.~\ref{fig:SimSpectrum} also shows
the translation motions, $\omega_{x}$, $\omega_{y}$ and $\omega_{z}$.

\section{Recoil heating}
We can estimate the recoil heating rates for translational and
rotational degrees of freedom due to gas collisions and photon
scattering using the Table.~\ref{table1}. These expressions can be
derived heuristically by considering the amount of linear and angular
momentum carried by a single gas particle or photon. Specifically, the
amount of linear and angular momentum carried by a gas particle can be
estimated as $\sqrt{2m_g k_b T}$ and $\sqrt{2m_g k_b T R^2}$,
respectively, where $m_g$ is the mass of a gas particle and $R$ is an
effective radius of the nanoparticle. To obtain the net effect on the
nanoparticle we have to take into account the gas collisions
scattering cross section: we formally replace $m_g$ with the mass of
the nanoparticle $M$ (we do not change the expressions in $\Gamma_c$,
which is associated with a single atom of the nanoparticle). We can
make a similar argument for photons: the amount of linear and angular
momentum carried by a photon particle is $\hbar \frac{2\pi}{\lambda}$
and $\hbar$, respectively. However, the photon scattering cross
section is already included in $\Gamma_s$, and thus we immediately
obtain the linear and angular momentum fluctuations per unit time in
Table.~\ref{table1}. A more rigorous derivation, based on the quantum
model~\cite{Toros2018t}, will be given elsewhere~\cite{Toros2018s}.

To get a rough numerical estimate we have considered the moment of inertia tensor of a sphere
and a susceptibility tensor that in the body frame has the diagonal
elements close to unity.
\begin{table}
\begin{tabular}{|c|c|c|}
  \cline{2-3} 
  \multicolumn{1}{c|}{} & gas collisions & photon scattering\tabularnewline
\hline 
translations & $ 4k_{B}TM\Gamma_{c}$  & $\Gamma_{s}\hbar^{2}\left(\frac{2\pi}{\lambda}\right)^{2}$\tabularnewline                                                                                             \hline 
rotations & $\frac{4}{5}k_{B}TMR^{2}\Gamma_{c}$ & $\Gamma_{s}\hbar^{2}$\tabularnewline
                                                                                                                                 \hline

\end{tabular}
\caption{The expressions denote estimates for the variance of momentum
  (angular momentum) fluctuations per unit time for translations
  (rotations), induced by gas collisions and photon scattering. Note
  that this expressions are good estimates for a system that is not
  highly anisotropic, while for a highly anisotropic objects, the
  moment of inertia and electric susceptibility tensors have to be
  taken into account~\cite{Toros2018s}. $M$ and $R$ denote an
  effective radius and mass of the nanoparticle, $\Gamma_{s}$ and
  $\Gamma_{c}$ denote photon scattering and gas collision rate,
  respectively, $T$ is the temperature of the gas, and $\lambda$ is
  the laser wavelength (see supplementary material A).}
\label{table1}
\end{table}

\begin{figure}[b]
  \includegraphics[width=0.5\textwidth]{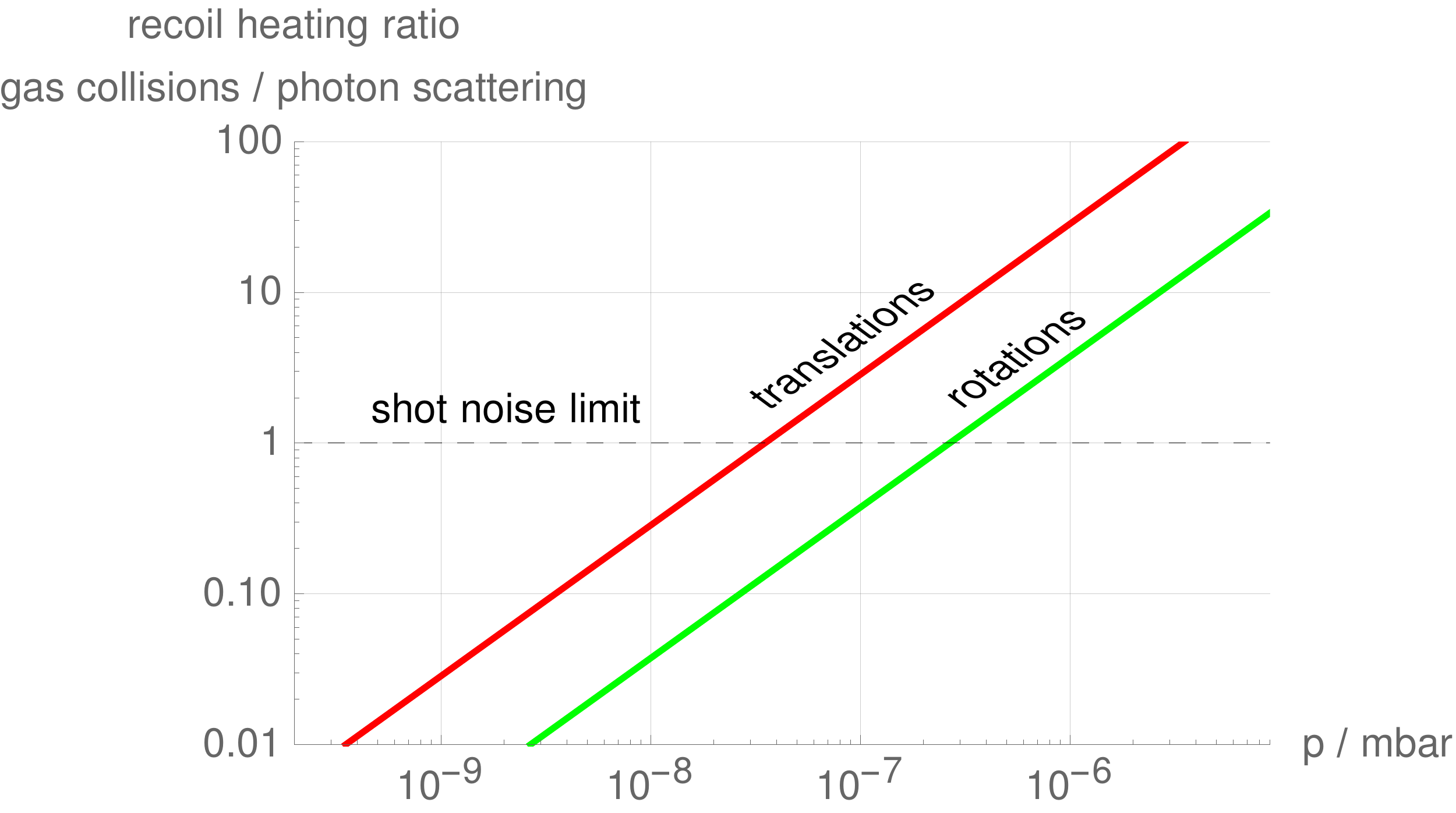}\caption{Comparison of
    recoil heating from gas collisions and from photon scattering.  We
    plot the ratio of the estimates in each row from
    Table.~\ref{table1}.  Values bigger (smaller) than
    1 mean that the recoil heating from gas collisions (photon
    scattering) is dominant. \label{fig:recoil_heating}}
\end{figure}

Using the formulae in Table.~\ref{table1} we compare
the strength of the heating mechanisms in Fig.~\ref{fig:recoil_heating}. This analysis also reproduces the transition to the photon recoil heating regime for translations is in accordance with Fig.~3 from \cite{vijay2016}.

\end{document}